\documentstyle[erice98,12pt,epsfig]{article}
\newcommand{\be}[1]{\begin{equation} \label{(#1)}}
\newcommand{\ee}{\end{equation}}
\newcommand{\ba}[1]{\begin{eqnarray} \label{(#1)}}
\newcommand{\ea}{\end{eqnarray}}

\newlength{\charwidth}

\def\vhight#1{\vphantom{\left(\begin{picture}(0,#1)\end{picture}\right)}
}
\def\Grho{\Green\thicklines
        \put(0.75,0){\oval(1.5,0.3)}
        \put(0,0){\Black\circle*{0.2}}\put(1.5,0){\Black\circle*{0.2}}
        \put(0,0){\line(1,0){0.8}}\put(1.5,0){\vector(-1,0){1}}}
\def\DSa{\begin{picture}(0,0)\thicklines
\put(0,0){\oval(1.5,1)}
\put(0,0){\makebox(0,0){$-\ii\Sa$}}\end{picture}}
\def\GlnG0Sa{
\begin{picture}(4.3,1.5)
\put(0.5,.1){
\put(.5,0){\DSa}\put(3,0){\DSa}\put(1.75,1){\DSa}
\put(1.75,-1){\makebox(0,0){\dots\dots}}
\put(1,.5){\oval(1,1)[lt]}\put(2.5,.5){\oval(1,1)[rt]}
\put(1,-.5){\oval(1,1)[lb]}\put(2.5,-.5){\oval(1,1)[rb]}}
\end{picture}}
\def\GGaSa{
\begin{picture}(2.5,2.)
\thicklines\put(0.5,.1){
\put(.5,0){\DSa}\put(2,-.5){\line(0,1){1}}
\put(1.,1){\line(1,0){0.5}}\put(1.,-1){\line(1,0){0.5}}
\put(1,.5){\oval(1,1)[lt]}\put(1.5,.5){\oval(1,1)[rt]}
\put(1,-.5){\oval(1,1)[lb]}\put(1.5,-.5){\oval(1,1)[rb]}}
\end{picture}}
\def\Dclosed#1#2{
\begin{picture}(2,.8)\put(0,.1){#2
\put(1,0){\circle{1.414}}
\put(1,-.707){\line(0,1){1.414}}\put(.293,0){\line(1,0){1.414}}
\put(0.5,-0.5){\line(0,1){1}}\put(1.5,-0.5){\line(0,1){1}}
\put(0.5,-0.5){\line(1,0){1}}\put(0.5,0.5){\line(1,0){1}}}
\put(1.5,-.8){$#1$}
\end{picture}}

\def\pls{\makebox(0,0){$+$}}
\def\mins{\makebox(0,0){$-$}}
\def\ssp{\makebox(0,0)
    {\thinlines\put(-.1,0){\line(1,0){.2}}\put(0,-.1){\line(0,0){.2}}}}
\def\ssm{\makebox(0,0){\put(-.1,0){\thinlines\line(1,0){.2}}}}
\def\photon{\thinlines\multiput(0,0)(.2,0){3}{\line(1,0){0.1}}}

\def\Photon{\thicklines\multiput(0,0)(0.2,0){5}{\line(1,0){0.1}}
\put(0.4,0){\vector(1,0){.2}}}

\def\Boson{\thicklines
           \multiput(0.0625,0)(.25,0){6}{\oval(.125,.125)[b]}
           \multiput(0.1875,0)(.25,0){6}{\oval(.125,.125)[t]}
           }

\def\VFermion{\thicklines\put(0,0){\vector(0,1){.8}}
            \put(0,.8){\line(0,1){.5}}}
\def\AVFermion{\thicklines\put(0,0){\vector(0,-1){.8}}
            \put(0,-.8){\line(0,-1){.5}}}

\def\oneloop{
     \put(1.5,0){\thicklines\oval(2.0,1.5)}
     \put(0,0){\photon}\put(0.3,0.3){\ssp} 
     \put(2.5,0){\photon}\put(2.7,0.3){\ssm}}

\def\oneloopvertex{
    \put(1.625,0){\thicklines\oval(2.0,1.5)}
    \put(0,0){\photon}\put(0.35,0.3){\ssp} 
    \put(0.625,0){\circle*{.25}}\put(2.625,0){\circle*{.25}}
    \put(2.75,0){\photon}\put(2.9,0.3){\ssm}}

\def\fullbox{\makebox(0,0){\rule{1.5mm}{3mm}}}
\def\interaction{\makebox(0,0){\put(0,0){\interact}
    \put(0,.95){\ssp}\put(0,-.95){\ssm}
    \put(0,.125){\ssp}\put(0,-.125){\ssm}}}
\def\interact{\makebox(0,0){\put(0,.5){\fullbox}
    \thicklines\put(0,0){\oval(.75,.5)}
    \put(0,-.5){\fullbox}} }

\def\til2loop{\put(0,0){\oneloopvertex}\put(4.,0){\pls}
      \put(4.75,0){\oneloopvertex}\put(6.375,0){\interaction}
      \put(9,0){\pls}}

\def\contourxy{
\begin{picture}(17,1)\thicklines
\contour
\put(14.5,-.5){\makebox(0,0){$\infty$}}
\put(11,-.5){\makebox(0,0){$t_x^+$}}
\put(8,1.8){\makebox(0,0){$t_y^-$}}
\put(11,0){\circle*{.2}}
\put(8,1){\circle*{.2}}
\end{picture}}

\def\contour{\thicklines
\put(1,-.5){\makebox(0,0){$t_{0}$}}
\put(16.5,.5){\makebox(0,0){$t$}}
\put(0,.5){\vector(1,0){16}}
\put(1,1.){\line(1,0){13}}\put(14,.5){\oval(1,1)[br]}
\put(14,0){\vector(-1,0){13}}\put(14,.5){\oval(1,1)[tr]}}

\newcommand{\di}{{\rm d}}
\newcommand{\ii }{{\rm i}}

\newcommand{\for}{\quad{\rm for}\quad}

\def\scr#1{\mbox{\scriptsize #1}}\def\Do{{\cal D}}\def\vu{v} 
\def\Lg{{\cal L}}
\def\Lgh{\makebox[3.5mm]{${\widehat{\makebox[2mm]{$\Lg$}}}$}\vphantom{L}}
\def\Lint{\Lgh^{\mbox{\scriptsize int}}}
\def\dpi#1{\frac{\di^4 #1}{(2\pi)^4}}                
\def\Pbr#1{\left\{#1\right\}}                    
\def\Gr{G}\def\Ga{G}\def\Se{\Sigma}\def\Sa{\Sigma}
\def\A{A}

\def\F{F}                             
\def\Ft{\widetilde{F}}                 
\def\Fd{F}                             
\def\Fdt{\widetilde{F}}                
\def\fd{f}

\def\Get{\Gamma_{\scr{out}}}   
\def\Gbt{\Get}                                   
\def\Ge{\Gamma_{\scr{in}}}     
\def\Gb{\Ge}              

\def\Ld{\Gamma_{\scr{out}}}
\def\Ldt{\Gamma_{\scr{in}}}
\def\Re{\mbox{Re}\;}\def\Im{\mbox{Im}\;}
\def\loopb#1{
\unitlength 1.00mm
\thicklines
\begin{picture}(17.50,13.00)
\put(2.00,1.00){\circle*{1.00}}
\put(17.00,1.00){\circle*{1.00}}
\bezier{68}(2.00,1.00)(9.50,5.00)(17.00,1.00)
\bezier{68}(2.00,1.00)(9.50,-3.00)(17.00,1.00)
\bezier{112}(2.00,1.00)(9.50,13.00)(17.00,1.00)
\bezier{112}(2.00,1.00)(9.50,-11.00)(17.00,1.00)
\put(9.93,-1.07){\vector(1,0){0.2}}
\put(9.00,-1.07){\line(1,0){0.93}}
\put(9.93,2.93){\vector(1,0){0.2}}
\put(8.93,2.93){\line(1,0){1.00}}
\put(9.13,6.93){\vector(-1,0){0.2}}
\put(10.07,6.93){\line(-1,0){0.93}}
\put(9.20,-5.00){\vector(-1,0){0.2}}
\put(10.07,-5.07){\line(-1,0){0.87}}
\if#1c \multiput(9.5,-8.)(0,6.5){3}{\thinlines\line(0,1){5}}\fi
\end{picture}}
\def\loopc#1{ 
\unitlength 1.00mm
\thicklines
\begin{picture}(18.54,9.47)
\put(2.00,-5.00){\circle*{0.93}}
\put(18.00,-5.00){\circle*{1.07}}
\put(10.00,9.00){\circle*{0.94}}
\bezier{72}(2.00,-5.00)(10.00,-1.00)(18.00,-5.00)
\bezier{72}(2.00,-5.00)(10.00,-9.00)(18.00,-5.00)
\bezier{80}(2.00,-5.00)(2.00,6.00)(10.00,9.00)
\bezier{72}(2.00,-5.00)(9.07,-0.80)(10.00,9.00)
\bezier{72}(18.00,-5.00)(11.07,-0.93)(10.00,9.00)
\bezier{76}(18.00,-5.00)(18.13,5.07)(10.00,9.00)
\put(9.93,-7.00){\vector(-1,0){0.2}}
\put(11.00,-7.00){\line(-1,0){1.07}}
\put(3.80,3.27){\vector(1,2){0.2}}
\multiput(3.40,2.47)(0.10,0.20){4}{\line(0,1){0.20}}
\put(16.73,2.33){\vector(2,-3){0.2}}
\multiput(16.20,3.13)(0.11,-0.16){5}{\line(0,-1){0.16}}
\put(10.07,-3.00){\vector(1,0){0.2}}
\put(9.07,-3.00){\line(1,0){1.00}}
\put(12.53,0.60){\vector(-2,3){0.2}}
\multiput(13.00,-0.20)(-0.12,0.20){4}{\line(0,1){0.20}}
\put(7.60,0.67){\vector(-2,-3){0.2}}
\multiput(8.00,1.33)(-0.10,-0.17){4}{\line(0,-1){0.17}}
\if#1c \multiput(8.5,-9.5)(3.5,5.25){3}{\thinlines\line(2,3){2.5}}\fi
\end{picture}}
\def\citerange[#1-#2]{[\citem[#1,@]-\citem[#2,@]]}
\def\citem[#1,#2]{\csname b@#1\endcsname\if @#2{}\else ,\citem[#2]\fi}
\let\section=\subsection     \let\subsection=\subsubsection                
\begin{document}
\title{
TRANSPORT DYNAMICS OF BROAD RESONANCES}

\author{J\"orn Knoll, GSI, Darmstadt}
\address{J.Knoll@gsi.de;
    http://theory.gsi.de} 
\date{Sept. 15, 1998}
\maketitle
\begin{abstract}
  The propagation of short life time particles with consequently broad
  mass width are discussed in the context of transport descriptions. In the
  first part some known properties of finite life time particles such as
  resonances are reviewed and discussed at the example of the $\rho$-meson.
  Grave deficiencies in some of the transport treatment of broad resonances
  are disclosed and quantified. The second part addresses the derivation of
  transport equations which permit to account for the damping width of the
  particles. Baym's $\Phi$-derivable method is used to derive a
  self-consistent and conserving scheme, which fulfils detailed balance
  relations even in the case of particles with broad mass distributions. For
  this scheme a conserved energy-momentum tensor can be constructed.
  Furthermore, a kinetic entropy can be derived which besides the standard
  quasi-particle part also includes contributions from fluctuations.
\end{abstract}
\section{Introduction and Prospects}
With the aim to describe the collision of two nuclei at intermediate or even
high energies one is confronted with the fact that the dynamics has to include
particles like the $\Delta^{33}$ or $\rho$-meson resonances with life-times of
less than 2 fm/c or equivalently with damping rates above 100 MeV. Also the
collision rates deduced from presently used transport codes are comparable in
magnitude, whereas typical mean kinetic energies as given by the temperature
range between 70 to 150 MeV depending on beam energy. Thus, the damping width
of most of the constituents in the system can no longer be treated as a
perturbation.

As a consequence the mass spectra of the particles in dense matter are no
longer sharp delta functions but rather acquire a width due to collisions and
decays. The corresponding quantum propagators $G$ (Green's functions) are no
longer the ones as in standard text books for fixed mass, but have to be
folded over a spectral function $A(\epsilon,{\vec p})$ of finite width. One
thus comes to a picture which unifies {\em resonances} which have already a
decay width in vacuum with the ``states'' of particles in dense
matter, which obtain a width due to collisions (collisional broadening). The
theoretical concepts for a proper many body description in terms of a real
time non equilibrium field theory have already been devised by Schwinger,
Kadanoff, Baym and Keldysh \cite{SKBK} in the early sixties. First
investigations of the quantum effects on the Boltzmann collision term were
given Danielewicz \cite{D}, the principal conceptual problems on the level of
quantum field theory were investigated by Landsmann \cite{Landsmann}, while
applications which seriously include the finite width of the particles in
transport descriptions were carried out only in recent times,
e.g.
 \citerange[D,DB-KV] For resonances, e.g. the $\Delta^{33}$-resonance, it was
natural to consider broad mass distributions and ad hoc recipes have been
invented to include this in transport simulation models.  However, many of
these recipes are not correct as they violate some basic principles like
detailed balance  \cite{DB}, and the description of resonances in dense matter
has to be improved.

In this talk the transport dynamics of short life time particles are reviewed
and discussed. In the first part some known properties of resonances are
presented. These concern the equilibrium and low density (virial) limits.
Some example discussions are given for the di-lepton spectrum resulting from
the decay of $\rho$-mesons in a dense nuclear environment, both in thermal
equilibrium and in a quasi-free scattering process. On the basis of this some
deficiencies of presently used transport codes for the treatment of broad
resonances are disclosed and quantified. They affect the di-lepton spectra
already on a qualitative level and signal that the low mass side is grossly
underestimated in the respective calculations. This motivates the question
discussed in the second part, namely, how to come to a self-consistent,
conserving and thermodynamically consistent transport description of particles
with finite mass width. The conceptual starting point will be a formulation
within the real-time non-equilibrium field theory.  The derivation is based on
and generalizes Baym's $\Phi$-functional method \cite{Baym}.  The first-order
gradient approximation provides a set of coupled equations of
time-irreversible generalized kinetic equations for the slowly varying
space-time part of the phase-space distributions supplemented by retarded
equations. The latter account for the fast micro-scale dynamics represented by
the four-momentum part of the distributions.  Functional methods permit to
derive a conserved energy-momentum tensor which also includes corrections
arising from fluctuations besides the standard quasi-particle terms. Memory
effects
 \citerange[CGreiner-IKV2] appearing in collision term diagrams of higher order
as well as the formulation of a non-equilibrium kinetic entropy flow can also
be addressed  \cite{IKV2}.

\section{Preliminaries}
The standard text-book transition rate in terms of Fermi's golden
rule, e.g. for the photon radiation from some initial
state $\left|i\right>$ with occupation $n_i$ to final states
$\left|f\right>$
\def\decay{\put(1,-1.1){\VFermion}\put(0.5,-.8){\makebox(0,0){$i$}}
\put(0.5,1.2){\makebox(0,0){$f$}}
\put(1,0.2){\VFermion}\put(1,0.2){\Photon}}
\def\decayA{\put(1.5,0.2){\AVFermion}\put(2,-.8){\makebox(0,0){$i$}}
\put(2,1.2){\makebox(0,0){$f$}}
\put(1.5,1.5){\AVFermion}\put(.5,.2){\Photon}}
\begin{eqnarray}\label{Wif}\unitlength5mm
W=&\sum_{if}&n_i(1-n_f)\;\unitlength5mm
\left|\begin{picture}(2.5,1.5)\decay\end{picture}\right|^{\;\mbox{2}}\;
(1+n_\omega)\;\delta(E_i-E_f-\omega_{\vec q})
\end{eqnarray}
with occupation $n_{\omega}$ for the photon, is limited to the concept of
asymptotic states. It is therefore inappropriate for problems which deal with
particles of finite life time. One rather has to go to the ``closed'' diagram
picture, where the same rate emerges as
\begin{eqnarray}\unitlength8mm\label{sigma-+}
W=\begin{picture}(3.75,1.3)\put(.2,.2){\oneloop}\thicklines
\put(1.9,.95){\vector(-1,0){.4}}\put(1.5,-.55){\vector(1,0){.4}}\end{picture}
(1+n_\omega)\delta(\omega-\omega_{\vec q})\\ \nonumber
\end{eqnarray}
with now two types of vertices $-$ and $+$ for the time-ordered and the
anti-time ordered parts of the square of the amplitude. Together with the
orientation of the $\stackrel{+~-}{\longrightarrow}$ and
$\stackrel{-~+}{\longrightarrow}$ propagator lines one obtains unique
diagrammatic rules for the calculation of rates rather than amplitudes. The
just mentioned propagator lines define the densities of occupied states or
those of available states, respectively. Therefore {\em all standard
diagrammatic rules} can be used again. One simply has to extend those rules to
the two types of vertices with marks $-$ and $+$ and the corresponding 4
propagators, the usual time-ordered propagator
$\stackrel{-~-}{\longrightarrow}$ between two $-$ vertices, the
anti-time-ordered one $\stackrel{+~+}{\longrightarrow}$ between two $+$
vertices and the mixed $\stackrel{+~-}{\longrightarrow}$ or
$\stackrel{-~+}{\longrightarrow}$ ones with fixed operator ordering
(Wightman-functions) as densities of occupied and available states. For
details I refer to the textbook of Lifshitz and Pitaevski \cite{LP}.
\unitlength10mm
\begin{center}
\unitlength10mm
\begin{picture}(17,3.)
\put(.0,1.){\contourxy}
\end{picture}\\
\small Fig.~1: Closed real-time contour with two external points $x,y$ on the
contour.
\end{center}

Equivalently the non-equilibrium theory can entirely be formulated on one
special time contour, the so called closed time path \cite{SKBK}, fig.~1, with
the time argument running from some initial time $t_0$ to infinity and back
with external points placed on this contour, e.g., for the four different
components of Green's functions or self energies. The special $-+$ or $+-$
components of the self energies define the gain and loss terms in transport
problems, c.f. eq. (\ref{sigma-+}) and eqs. (\ref{Coll(kin)}-\ref{G-def})
below.

The advantage of the formulation in terms of ``correlation'' diagrams, which
no longer refer to amplitudes but directly to physical observables,
like rates, is that now one is no longer restricted to the concept of
asymptotic states. Rather all internal lines, also the ones which originally
referred to the ``in'' or ``out'' states are now treated on equal footing.
Therefore now one can deal with ``states'' which have a broad mass spectrum.
The corresponding Wigner densities $\stackrel{+~-}{\longrightarrow}$ or
$\stackrel{-~+}{\longrightarrow}$ are then no longer on-shell
$\delta$-functions in energy (on-mass shell) but rather acquire a width, as
we shall discuss in more detail.

For slightly inhomogeneous and slowly evolving systems, the degrees of freedom
can be subdivided into rapid and slow ones. Any kinetic approximation is
essentially based on this assumption.  Then for any two-point function
$F(x,y)$, one separates the variable $\xi =(t_1-t_2, \vec{r_1}-\vec{r_2})$,
which relates to the rapid and short-ranged microscopic processes, and the
variable $X= \frac{1}{2}(t_1+t_2,\vec{r_1}+\vec{r_2})$, which refers to slow
and long-ranged collective motions. The Wigner transformation, i.e.  the
Fourier transformation in four-space difference $\xi=x-y$ to four-momentum $p$
of the contour decomposed components of any two-point contour function
%
\begin{equation}
\label{W-transf} 
F^{ij}(X;p)=\int \di \xi e^{\ii p\xi}
F^{ij}\left(X+\xi/2,X-\xi/2\right),\quad\mbox{where $i,j\in\{-+\}$ }
\end{equation}
%
leads to a (co-variant) four phase-space formulation of two-point functions.
The Wigner transformation of Dyson's equation (\ref{varG/phi}) in $\{-+\}$
notation is straight forward. For details and the extensions to include the
coupling to classical field equations we refer to ref. \cite{IKV1}.

Standard transport descriptions usually involve two approximation steps: (i)
the gradient expansion for the slow degrees of freedom, as well as (ii) the
quasi-particle approximation for rapid ones. We intend to avoid the latter
approximation and will solely deal with the gradient approximation for slow
collective motions by performing the gradient expansion of the coupled Dyson
equations. This step indeed preserves all the invariances of the $\Phi$
functional in a $\Phi$-derivable approximation.

It is helpful to avoid all the imaginary factors inherent in the standard
Green's function formulation and change to quantities which are real and
positive either in the homogeneous or in appropriate coarse graining
limits. They then have a straight physical interpretation analogously to the
Boltzmann equation.  We define
%
\begin{eqnarray}
\label{F}
\left.
\begin{array}{rcl}
\Fd (X,p) &=& \A (X,p) \fd (X,p)
 = \mp \ii \Gr^{-+} (X,p) ,\\
\Fdt (X,p) &=& \A (X,p) [1 \mp \fd (X,p)] = \ii \Gr^{+-} (X,p),
\end{array}\right\}\quad{\rm with}\quad
\label{A}
 A (X,p) &\equiv& -2\Im \Gr^R (X,p) = \Fdt \pm \Fd\hspace*{0.5cm}
\end{eqnarray}
%
for the generalized Wigner functions $\F$ and $\Ft$ with the corresponding {\em
  four} phase space distribution functions $\fd(X,p)$, the  Fermi/Bose factors
  $[1 \mp \fd (X,p)]$ and spectral function $A (X,p)$. According to the
  retarded relations between Green's functions $\Gr^{ij}$, {\em only two of
  these real functions are required for a complete dynamical
  description}. Here and below upper signs relate to fermion quantities,
  whereas lower signs refer to boson quantities. As shown in ref. \cite{IKV1}
  mean fields and condensates, i.e. non-vanishing expectation values of
  one-point functions can also be included.

\section{Thermodynamic Equilibrium}

The thermodynamic equilibrium leads to a lot of simplifying relations among
the kinetic quantities. All quantities become space-time independent. The
Kubo-Martin-Schwinger condition determines the distribution functions to be of 
Fermi-Dirac or Bose-Einstein type, respectively
%
\begin{eqnarray}
\label{Feq}
f_{\rm eq}(X,p)=1/\left(\exp\left((p_0-\mu)/T\right)\pm 1\right).
\end{eqnarray}
%
Here $\mu$ is the chemical potential. The spectral function attains a form
%
\begin{eqnarray}
\label{Aeq}
A_{\rm eq}(X,p)&=&\frac{\Gamma(p)}{M^2(p)+\Gamma^2(p)/4}\quad{\rm with}\quad
\left\{
\begin{array}{rcl}
\Gamma(p)&=&-2\Im \Sa^{\rm R}(p),\\ 
M(p)&=&M_0(p)-\Re \Sa^{\rm R}(p),\quad M_0(p)=p_0^{\kappa}-p_0^{\kappa}({\vec
  p}). 
\end{array}\right. 
\end{eqnarray}
%
This form is exact through the four-momentum $p=(p_0,{\vec p})$ dependence of
the retarded self-energy $\Sa^{\rm R}(p)$. Thereby
$M_0(p)=p_0^{\kappa}-p_0^{\kappa}({\vec p})=0$ is the free dispersion relation
with $\kappa=1$ or 2 for the non-relativistic Schr\"odinger or the
relativistic Klein-Gordon case, respectively. In the non-equilibrium case all
quantities become functions of the space-time coordinates $X$ and, of course,
the distribution functions $f(X,p)$ generally also depend on three momentum
$\vec p$.

\section{The Virial Limit}
Another simplifying case is provided by the low density limit, i.e. the virial
limit. Since Beth-Uhlenbeck (1937) \cite{BethU} it is known that the
corrections to the level density are given by the asymptotic properties of
binary scattering processes, i.e. in a partial wave decomposition by means of
phase-shifts, see also 
\citerange[Huang-Mekjian]
. The reasoning can be
summarized as follows. While for any pair the c.m. motion remains unaltered
the relative motion is affected by the mutual two-body interaction.
Considering a large quantization volume of radius $R$ and a partial wave of
angular momentum $j$, the levels follow the
quantization condition \def\Blue{}\def\blue{}
\def\Red{}\def\Green{}\def\Black{}
\begin{eqnarray}\label{psi}
\unitlength1cm
\begin{picture}(8,1.5)
\put(0,-0.6){
\put(0,1.7){$\Blue\psi_j(r)\longrightarrow \sin(kr+{\Red \delta_j(E)})$}
\put(7.93,0.8){\makebox(-0.2,0){$|$}}
\put(8.1,0.4){\makebox(-0.2,0){$R$}}
\put(0,0){\epsfig{file=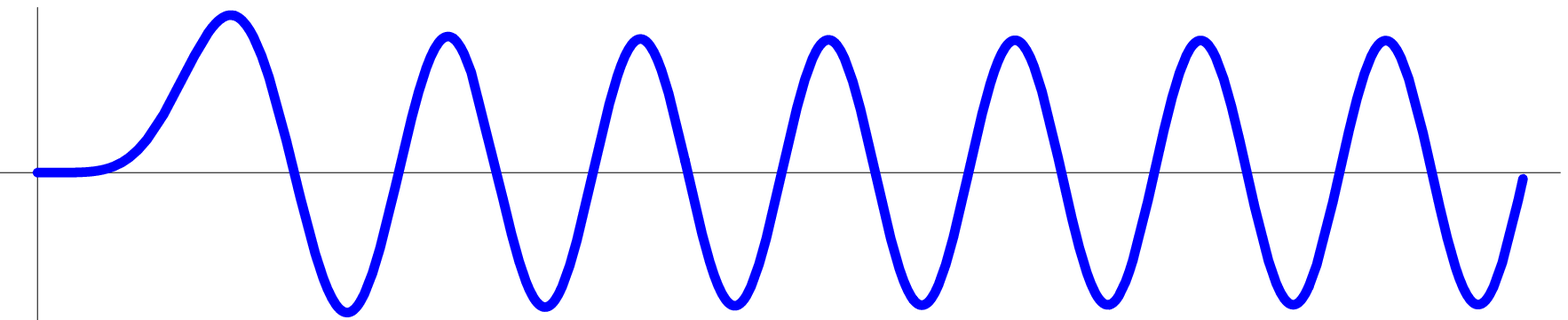,width=8cm,height=1.5cm}}
}
\end{picture}\quad\quad\Rightarrow\quad\quad
kR+\delta_j(E)=n\pi,
\\ \nonumber
\end{eqnarray}
where $\delta_j(E)$ is the phase-shift at relative energy $E$ and $n$ is an
integer counting the levels. The $kR$ term accounts for the free motion part.
The corresponding corrections to both, the level density and thermodynamic
partition sum, are
given by\\
\begin{minipage}[b]{7cm}
\begin{picture}(6,10.5)
  \put(0.5,3.4){
        \put(0,0){\epsfig{file=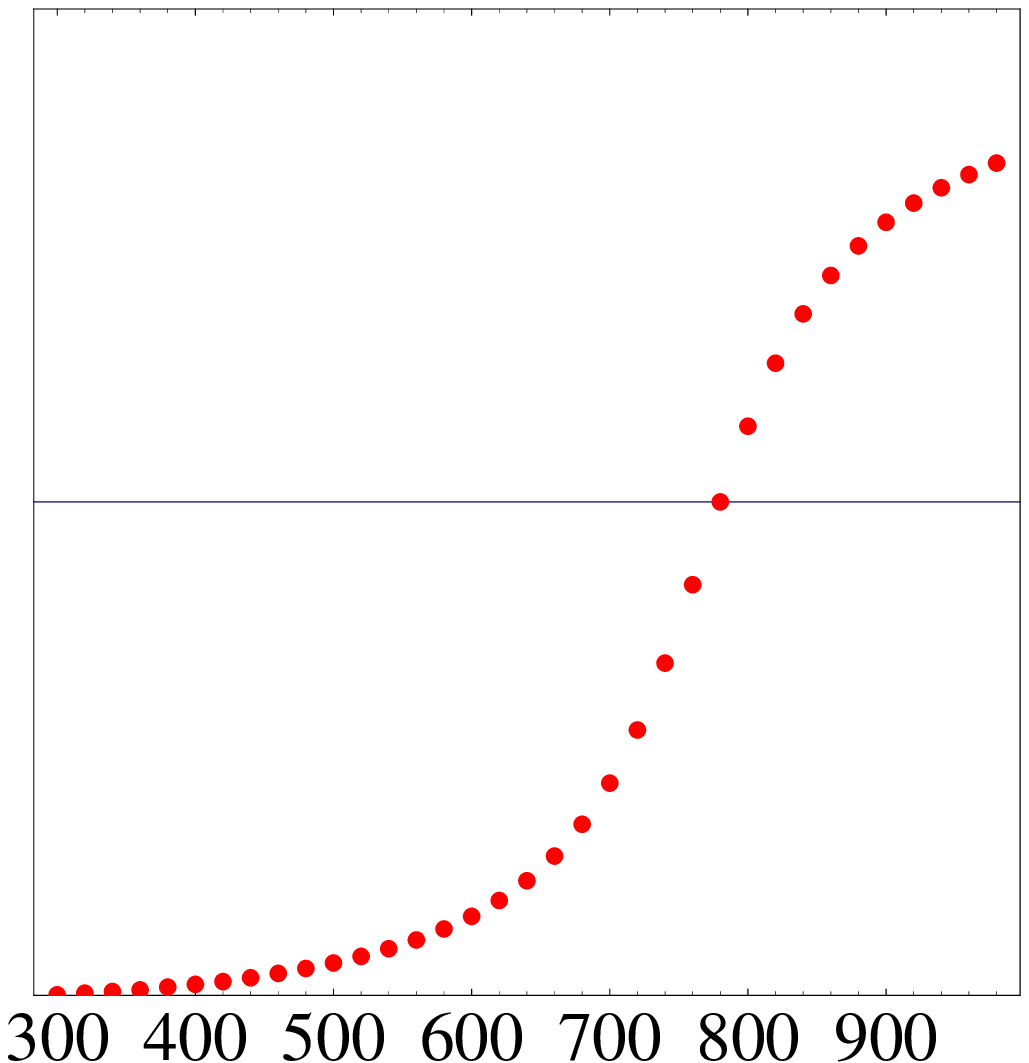,width=6cm,height=7cm}}
        \put(3,5.7){\makebox(0,0){$\Red\delta(\pi^+\pi^-)$}}
        \put(0,6.95){\makebox(0,0){$\pi$}}
        \put(0,3.678){\makebox(0,0){$\displaystyle\frac{\pi}{\small 2}$}}
        \put(3,-.3){\makebox(0,0){\small$E\mbox{ [MeV]}$}}}
  \put(1.4,1.2){\Blue
        \put(1.5,0){\vector(-1,1){0.8}}\put(0.4,1.1){\line(1,-1){0.8}}
        \put(1.5,0){\vector(-1,-1){0.8}}\put(0.4,-1.1){\line(1,1){0.8}}
        \put(0,1.2){\makebox(0,0){$\pi^+$}}
        \put(0,-1.1){\makebox(0,0){$\pi^-$}}
        \put(1.5,0){\Grho}\put(2.25,.7){\makebox(0,0){\Green$\rho$}}
        \put(3,0){\line(1,1){0.8}}\put(4.1,1.1){\vector(-1,-1){0.6}}
        \put(3,0){\line(1,-1){0.8}}\put(4.1,-1.1){\vector(-1,1){0.6}}
        \put(4.5,1.2){\makebox(0,0){\small $\pi^+$}}
        \put(4.5,-1.1){\makebox(0,0){\small $\pi^-$}}
        \put(1.5,0){\circle*{0.2}}
        \put(3,0){\circle*{0.2}}
        }
                \end{picture}
\begin{center}\small Fig.~2: $\pi^+\pi^-$ $p$-wave phase-shifts\\ 
and scattering diagram.\\[3mm] $ $
\end{center}
\end{minipage}
\begin{minipage}[b]{11.2cm}
\begin{eqnarray}
\displaystyle
        {\Red\frac{\di n}{\di E}}&=&\frac{\di n^{\rm free}}{\di E}
        +{\frac{2j+1}{\pi}\;\frac{\di {\Red\delta_j}}{\di E}}\\[3mm]
        Z&=&\sum_i e^{-E_i/T}=\int \di E {\Red\frac{\di n}{\di E}} e^{-E/T}
\end{eqnarray}
Since $Z$ determines the equation of state, its low density limit is uniquely
given by the scattering phase-shifts.  The energy derivatives of the
phase-shifts are also responsible for the time-delays discussed in ref.
\cite{DP} and also for the virial corrections to the Boltzmann collision term
recently discussed in ref.  \cite{Mor98}. The latter is directly connected to
the $B$-term in our generalized kinetic equation (\ref{keqk}).  The advance of
a phase-shift by a value range of $\pi$ across a certain energy window adds
one state to the level density and points towards an $s$-channel resonance. An
example is the $\rho$-meson in the $p$-wave $\pi^+\pi^-$ scattering channel,
fig.~2. In cases, where the resonance couples to one asymptotic channel only,
the corresponding phase-shifts relate to the vacuum spectral function $A_j(p)$
of that resonance via\footnotemark
\begin{eqnarray}\label{Tinout}
4\left|T_{\rm in,out} \right|^2&=&
    \frac{
        {\Green\Gamma_{\Blue\rm in}(E)\Green\Gamma_{\Blue\rm out}(E)}}
        {\left(E^{\kappa}-{\Green E_R^{\kappa}(E)}\right)^2
          +{\Green \Gamma_{\Red\rm tot}^2(E)}/4}
        \\[0.5cm]\label{Tsingle}
        &=&4\;\sin^2{\Blue \delta_j(E)\;
          =\Green A_j(E,{\vec p}=0)\;\Gamma_{\Blue\rm tot}(E)}.\\[-2mm]\nonumber
\end{eqnarray}
\end{minipage}\\
\noindent
Here $T_{\rm in,out}$ is the corresponding $T$-matrix element.  While relation
(\ref{Tinout}) is correct also in the case where many channels couple to the
same resonance, relation (\ref{Tsingle}) only holds for the single channel
case, where $\Gamma_{\rm in}=\Gamma_{\rm out}=\Gamma_{\rm tot}$.  Relation
(\ref{Tsingle}) illustrates that the vacuum spectral functions of resonances
can almost model-independently be deduced from phase-shift information. In the
case of the $\rho$-meson additional information is provided by the pion form
factor.  Also in the case of two channels coupling to a resonance the energy
dependence of phase-shifts of the two scattering channels together with the
inelasticity coefficient provide stringent constraints for the spectral
function of the resonance \cite{Weinh-PhD}.  \footnotetext{$E$ is the relative
  c.m. energy and correspondingly the momentum in $A$ vanishes; $\kappa=1$ for
  {non-rel.} particles; $\kappa=2$ for relativistic bosons, where
  $\Gamma(E)/2E$ equals the energy dependent decay width and
  \mbox{$E_R^2(E)=m_R^2+{\vec p}^2+\Re\Se(p)$.}}

\section{The $\rho$-meson in dense matter}

An an example I like to discuss the properties of the $\rho$-meson and the
consequences for the decay into di-leptons.  The exact production rate of
di-leptons is given by the following formula \unitlength8mm
\begin{eqnarray}\label{dndtdm}
\frac{\di n^{\mbox{e}^+\mbox{e}^-}}{\di t\di m}&=&
\begin{picture}(7.5,1.5)\put(0.,0.2){
        \put(1.5,0){\vector(-1,1){0.8}}\put(0.2,1.3){\line(1,-1){0.55}}
        \put(1.5,0){\vector(-1,-1){0.8}}\put(0.2,-1.3){\line(1,1){0.55}}
        \put(.8,1.2){\makebox(0,0){\small e$^+$}}
        \put(.8,-1.2){\makebox(0,0){\small e$^-$}}
        \put(1.5,0){\Boson}\put(2.25,-.7){\makebox(0,0){$\gamma^*$}}
        \put(3,0){\Grho}\put(3.75,-.7){\makebox(0,0){\Green$\rho$}}
        \put(3,0.5){\mins}\put(4.5,0.5){\pls}
        \put(4.5,0){\Boson}\put(5.25,-.7){\makebox(0,0){$\gamma^*$}}
        \put(6,0){\line(1,1){0.8}}\put(7.3,1.3){\vector(-1,-1){0.6}}
        \put(6,0){\line(1,-1){0.8}}\put(7.3,-1.3){\vector(-1,1){0.6}}
        \put(6.8,1.2){\makebox(0,0){\small e$^+$}}
        \put(6.8,-1.2){\makebox(0,0){\small e$^-$}}
        \put(1.5,0){\circle*{0.2}}
        \put(6,0){\circle*{0.2}}
        }
\end{picture}
={\Green f_{\rho}(m,{\vec p},{\vec x},t)\;
        A_{\rho}(m,{\vec p},{\vec x},t)}\;
\Gamma^{\rho\;\mbox{\small e}^+\mbox{\small e}^-}(m).
\vphantom{\left(\begin{picture}(0,1.5)\end{picture}\right)}
\end{eqnarray}
Here $\Gamma^{\rho\;\mbox{\small e}^+\mbox{\small e}^-}(m)\propto1/m^2$ is the
mass-dependent electromagnetic decay rate of the $\rho$-meson into the
di-electron channel. The phase-space distribution $f_{\rho}(m,{\vec p},{\vec
  x},t)$ and the spectral function $A_{\rho}(m,{\vec p},{\vec x},t)$ define
the properties of the $\rho$-meson at space-time point ${\vec x},t$. Both
quantities are in principle to be determined dynamically by an appropriate
transport model. However till to-date the spectral functions are not treated
dynamically in most of the present transport models. Rather one employs
on-shell $\delta$-functions for all stable particles and spectral functions
fixed to the vacuum shape for resonances.

As an illustration the model case is discussed, where the $\rho$-meson just
strongly couples to two channels: naturally the $\pi^+\pi^-$ channel and to
the $\pi N\leftrightarrow\rho N$ channels relevant at finite nuclear
densities. The latter component is representative for all channels
contributing to the so-called {\em direct $\rho$} in transport codes. For a
first orientation the equilibrium properties are discussed.  Admittedly by far
more sophisticated and in parts unitary consistent equilibrium calculations
have already be presented in the literature, e.g.
\citerange[Mosel-FLW].  It is not the point to compete with them at this
place. Rather we try to give a detailed analysis in simple terms with the aim
to discuss the consequences for the implementation of such resonance processes
into dynamical transport simulation codes.

Both considered processes add to the total width of the $\rho$-meson
\begin{eqnarray}\label{Gammatot}
\Gamma_{\rm tot}(m,{\vec p})&=&\Gamma_{\rho\rightarrow{\pi}^+{\pi}^-}(m,{\vec
  p})+ 
\Gamma_{\rho\rightarrow{\pi} NN^{-1}}(m,{\vec p}),
\end{eqnarray}
and the equilibrium spectral function then results from the cuts of the
two diagrams 
\unitlength6mm
\begin{eqnarray}\label{A2}
{\Green A_{\rho}(m,{\vec p})}\;&=&\normalsize
\begin{picture}(5.5,1.3)\thicklines\put(0.25,0.2){
        \put(0,0){\Grho}
        \put(3.5,0){\Grho}
        \put(2.5,0){\Blue\oval(2,1.5)}
        \put(2.6,0.75){\Blue\vector(-1,0){0.3}}
        \put(2.6,-0.75){\Blue\vector(-1,0){0.3}}
        \put(2.5,1.25){\makebox(0,0){\Blue$\pi^+$}}
        \put(2.5,-1.25){\makebox(0,0){\Blue$\pi^-$}}
        \put(1.95,-1.1){\thinlines\line(1,2){0.3}}
        \put(2.35,-0.3){\thinlines\line(1,2){0.3}}
        \put(2.75,.5){\thinlines\line(1,2){0.3}}
        }
\end{picture} +
\begin{picture}(5.4,1)\thicklines\put(0.25,0.2){
        \put(0,0){\Grho}
        \put(3.5,0){\Grho}
        \put(2.5,0){\blue\oval(2,1.5)[b]}
        \put(2.5,0){\Red\oval(2,1.5)[t]}
        \put(2.4,0.75){\Red\vector(1,0){0.3}}
        \put(2.6,-0.75){\Blue\vector(-1,0){0.3}}
        \put(2.5,1.25){\makebox(0,0){\Red$ N^{-1}$}}
        \put(2.45,0.35){\makebox(0,0){\Blue$\pi$}}
        \put(2.5,-1.25){\makebox(0,0){\Blue N}}
        \put(3.5,0){\Blue\vector(-1,0){1.3}}
        \put(1.5,0){\Blue\line(1,0){1}}
        \put(0.05,0){
          \put(1.95,-1.1){\thinlines\line(1,2){0.3}}
          \put(2.35,-0.3){\thinlines\line(1,2){0.3}}
          \put(2.75,.5){\thinlines\line(1,2){0.3}}}
        }
\end{picture}
=\frac{
        {\Blue\Gamma_{\rho\;\pi^+\pi^-}} +
        {\Red\Gamma_{\rho\;\pi N N^{-1}}}}
        {\left(m^2-m_\rho^2-\mbox{Re}\Sigma\right)^2
        +{\Red \Gamma_{\rm tot}^2/4}}\; .
\end{eqnarray}
In principle both diagrams have to be calculated by fully self consistent
propagators, i.e. with corresponding widths for all particles involved. This
formidable task has not been done yet. Using micro-reversibility and the
properties of thermal distributions the two terms in (\ref{A2}) contributing
to the di-lepton yield (\ref{dndtdm}) can indeed approximately be reformulated
as the thermal average of a $\pi^+\pi^-\rightarrow\rho\rightarrow{\rm e}^+{\rm
e}^-$-annihilation process and a $\pi N\rightarrow\rho N\rightarrow{\rm
e}^+{\rm e}^-N$-scattering process, i.e.
\begin{eqnarray}\label{x-sect}
\frac{\di n^{{\rm
  e}^+{\rm e}^-}}{\di m\di t}\propto
    \left<f_{\pi^+}f_{\pi^-}\; v_{\pi\pi}\;
      \sigma(\pi^+\pi^-\rightarrow\rho\rightarrow{\rm e}^+{\rm
          e}^-)\vphantom{A^A}+
      f_{\pi}f_N\; v_{\pi N}\;\sigma(\pi N\rightarrow\rho N\rightarrow{\rm
          e}^+{\rm e}^-N)\vphantom{A^A}\right>_T
\end{eqnarray}
However, the important fact to be noticed is that in order to preserve
unitarity the corresponding cross sections are no longer the free ones, as
given by the vacuum decay width in the denominator, but rather involve the
{\em medium dependent total width} (\ref{Gammatot}). This illustrates in
simple terms that rates of broad resonances can no longer simply be added in a
perturbative way.  Since it concerns a coupled channel problem there is a
cross talk between the different channels to the extent that the common
resonance propagator attains the total width arising from all partial widths
feeding and depopulating the resonance. While a perturbative treatment with
free cross sections in (\ref{x-sect}) would enhance the yield at resonance,
$m=m_{\rho}$, if a channel is added, c.f. fig.~2 left part, the correct
treatment (\ref{A2}) even inverts the trend and indeed depletes the yield at
resonance, right part in fig.~2. Furthermore one sees that only the total yield
involves the spectral function, while any partial cross section only refers to
that partial term with the corresponding partial width in the numerator!
Unfortunately so far all these facts have been ignored or even overlooked in
the present transport treatment of broad resonances. 

\noindent
\unitlength1cm
\begin{picture}(19,6.3)
\put(0,0){\epsfig{file=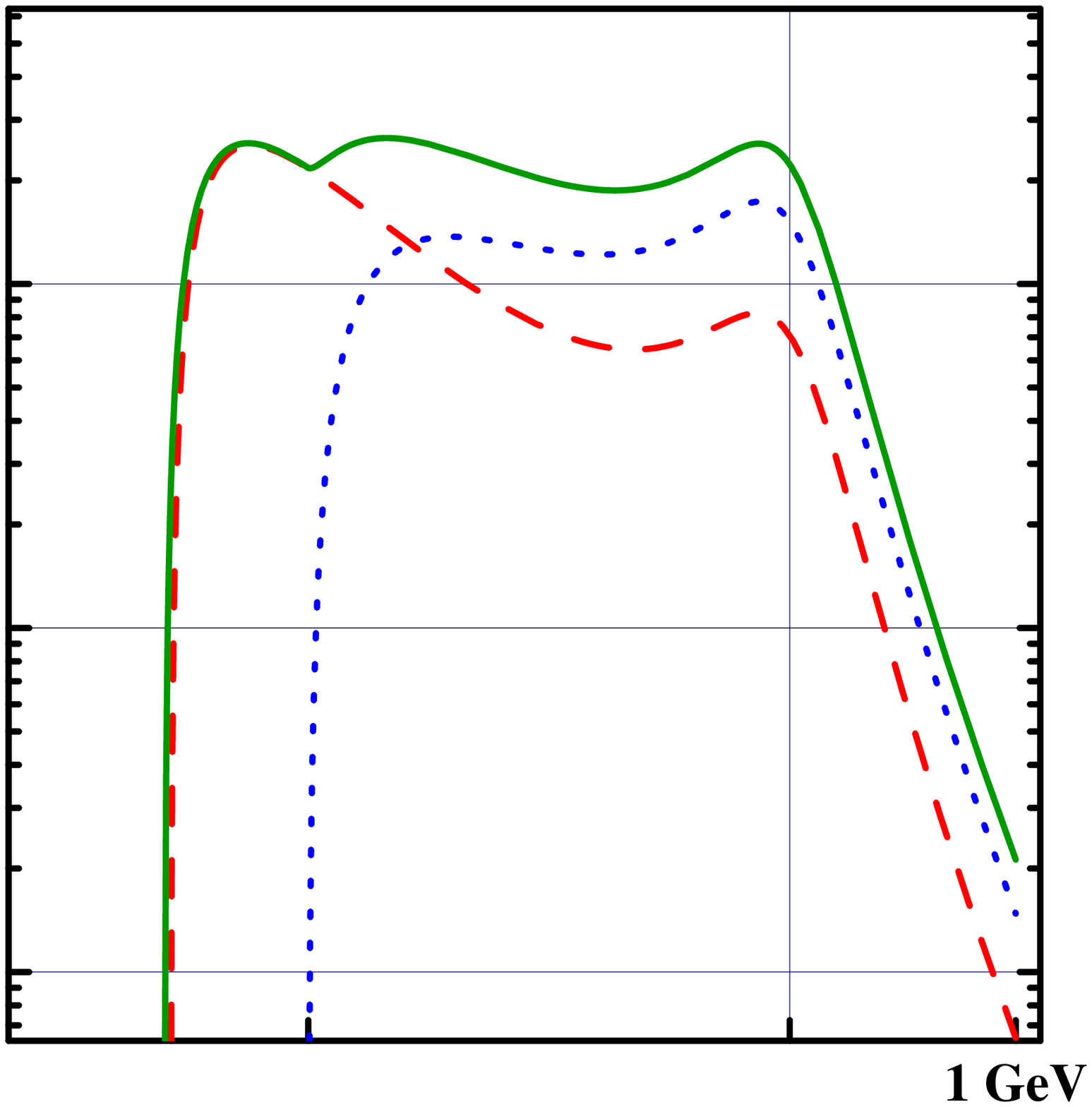,width=6cm,height=5.7cm}}
\put(6.,0){\epsfig{file=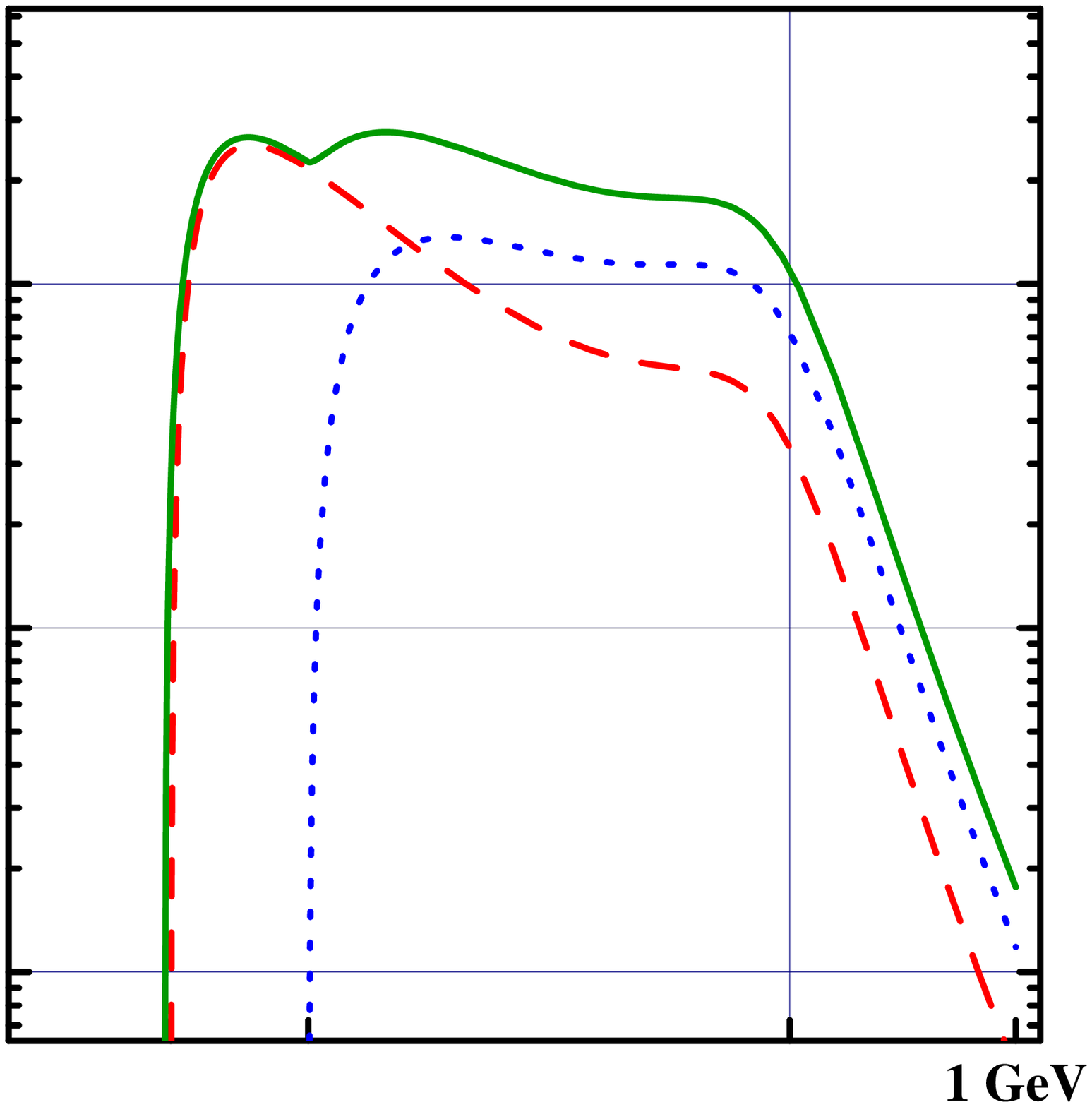,width=6cm,height=5.7cm}}
\put(11.9,0.08){\epsfig{file=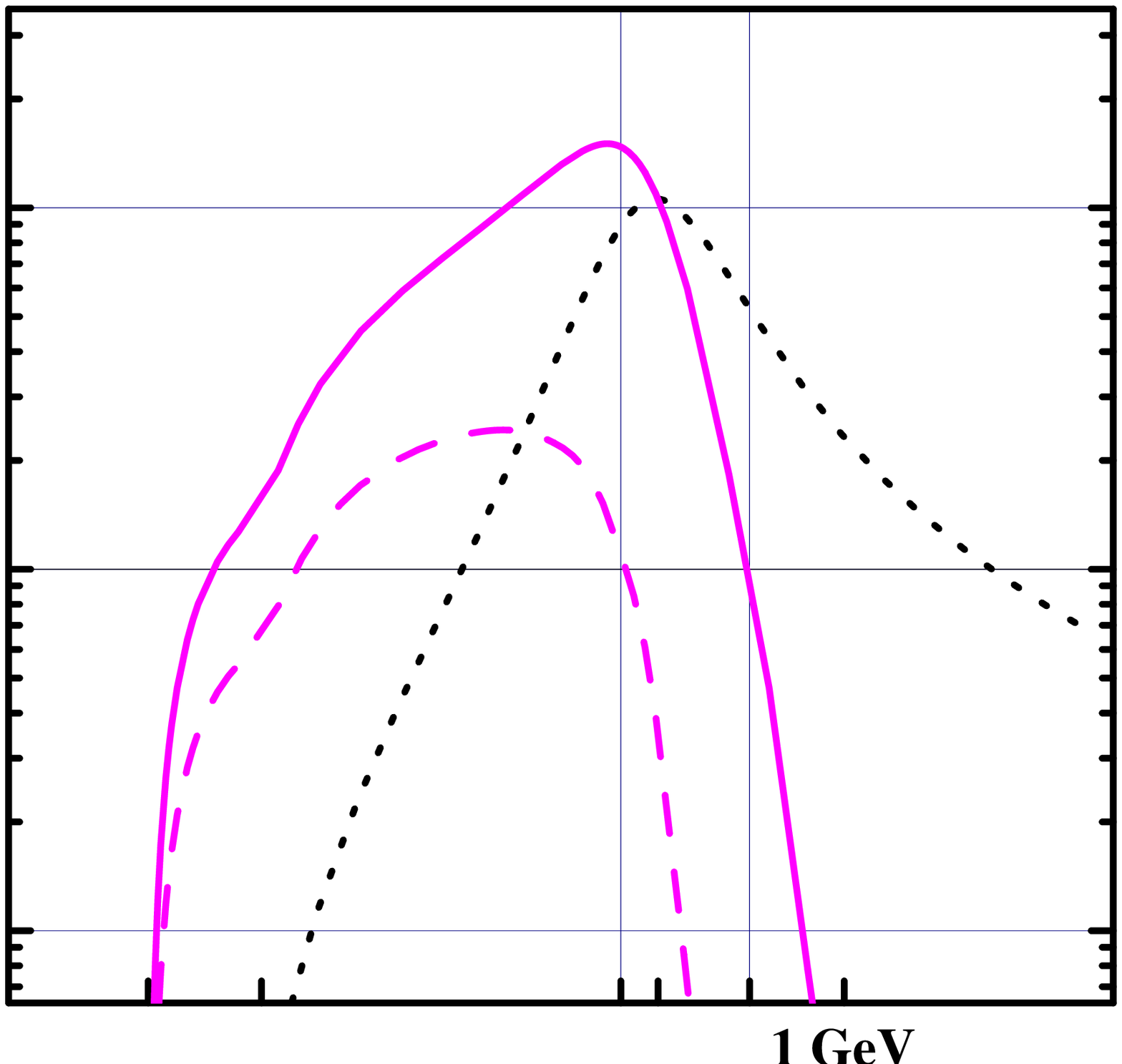,width=6.5cm,height=5.2cm}}
\put(6.,6.){\makebox(0,0){\bf Di-lepton rates from thermal $\rho$-mesons
($T=110$ MeV)}}
\put(15.,6.){\makebox(0,0){\bf Quasi-free $\pi N$ collisions}}
\put(15.,5.6){\makebox(0,0){\bf and spectral function}}
\put(1.,0.1){\makebox(0,0){\small$m_{\pi}$}}
\put(1.8,0.1){\makebox(0,0){\small$2m_{\pi}$}}
\put(4.3,0.1){\makebox(0,0){\small$m_{\rho}$}}
\put(6,0){
\put(1.,0.1){\makebox(0,0){\small$m_{\pi}$}}
\put(1.8,0.1){\makebox(0,0){\small$2m_{\pi}$}}
\put(4.3,0.1){\makebox(0,0){\small$m_{\rho}$}}}
\put(3,5.3){\makebox(0,0){$\Gamma_{\rm tot}=\Gamma_{\rm free}$}}
\put(9,5.3){\makebox(0,0){full $\Gamma_{\rm tot}$}}
\put(12,0){
\put(1.6,0.1){\makebox(0,0){\small$2m_{\pi}$}}
\put(3.8,0.1){\makebox(0,0){\small$m_{\rho}$}}}
\end{picture}
\parbox[t]{11.7cm}{\small Fig.~3: $\mbox{e}^+\mbox{e}^-$ rates (arb.  units)
  as a function of the invariant pair mass $m$ at $T=110$ MeV from
  $\pi^+\pi^-$ annihilation (dotted line) and direct $\rho$-meson contribution
  (dashed line), the full line gives the sum of both contributions. Left part:
  using the free cross section recipe, i.e. with $\Gamma_{\rm
    tot}=\Gamma_{\rho\;\pi^+\pi^-}$; right part for the correct partial rates
  (\ref{A2}). The calculation are done with
  $\Gamma_{\rho\leftrightarrow\pi\pi}(m_{\rho})/2m_{\rho}=150$ MeV and
  $\Gamma_{\rho\leftrightarrow\pi N N^{-1}}(m_{\rho})/2m_{\rho}=70$
  MeV.\\}\hfill \parbox[t]{6.2cm}{\small Fig.~4: Fermi motion averaged $\pi
  N\rightarrow\rho N\rightarrow {\rm e}^+{\rm e}^- N$ cross sections at pion
  beam momenta of 1 and 1.3 GeV/c (dashed and full curve) as a function of
  invariant pair mass $m$. The dotted line
  gives the spectral function used here and in fig.~3.\\}

Compared to the spectral function (dotted line in fig.~4) both thermal
components in fig.~3 show a significant enhancement on the low mass side and a
strong depletion at high masses due to the thermal weight
$f\propto\exp(-p_0/T)$ in the rate (\ref{dndtdm}).  A similar effect is seen
in genuine non-equilibrium processes like the di-lepton yield resulting from
Fermi-motion averaged $\pi N\rightarrow \rho N$ scattering, fig.~4.  The
latter is representive for the first chance collision in a $\pi A$ reaction
and shows a behavior significantly different from that obtained in
refs.~\cite{CassingpiA}. For orientation the sub-threshold conditions for the
two beam momenta are given by the vertical lines in fig.~4!

Much of the physics can already be discussed by observing that the partial
widths are essentially given by the type of coupling ($s$ or $p$-wave,
$l=0,1$) and the phase space available for the decay channel. For point-like
couplings and two-body phase space or approximately in the case of one light
particle among only heavy ones in the decay channel (e.g. for $\pi N N^{-1}$)
one finds
\begin{eqnarray}\label{Gamma(m)}
\Gamma_{c}(m)\propto m p_{\rm cm}\left(\frac{p_{\rm cm}}{m}\right)^l
\quad\mbox{with}\quad p_{\rm cm}\propto\sqrt{
m^2-s_{\rm thr}} ;\;
\left\{ 
\begin{array}{lll}
        s_{\rm thr}=4m_{\pi}^2,&\quad l=1\;
        &\mbox{for } c=\left\{\rho\leftrightarrow\pi\pi\right\}\\
        s_{\rm thr}=m_{\pi}^2,&\quad l=1
        &\mbox{for } c=\left\{\rho\leftrightarrow\pi N N^{-1}\right\}.
\end{array}\right.\hspace*{4mm} 
\end{eqnarray}
In the $\pi\pi$ case the corresponding strength is approximately given by the
vacuum decay, while it depends on the nuclear density in the $\rho
N\leftrightarrow\pi N$ case.  The simple phase-space behavior (\ref{Gamma(m)})
suggests that far away from all thresholds ($m^2\gg s_{\rm thr}$) the ratio
$\left<\pi\pi\mbox{-annihilation}\right>/\left<\mbox{direct }\rho\right>$ of
the two components should result to a fairly {\em smooth} and {\em almost
constant} function of $m$, e.g. for $m>500$ MeV. This kinematical constraint
is nicely confirmed by some calculations, e.g. of ref. \cite{Ko}, however, no
such behavior is seen in the {\em direct} $\rho$-mesons so far computed in
refs. \cite{CassingpiA,Cassing}\footnote{In refs.  \cite{CassingpiA,Cassing}
the direct $\rho$ component appears almost like the spectral function itself,
i.e. untouched from any phase-space constraints which come in through the
distributions $f_{\rho}(X,p)$. The latter favour low masses and deplete the
high mass components! In fact rather than being almost constant the ratios
$\left<\pi\pi\right>/\left<\mbox{direct }\rho\right>$ exhibit an exponential
behavior $\exp(-m/T^*)$ for $m>500$ MeV with $T^*$ between 70 - 110 MeV
depending on beam energy, pointing towards a major deficiency in the account
of phase-space constraints for the {\em direct} $\rho$-meson component in
these calculations.}.

For completeness and as a stimulus for improvements the discussed  defects
in some of presently used transport treatments of broad resonances (vector
mesons in particular) are listed below. The last column gives an estimate of
multiplicative factors needed to restore the defect ($m_R$ is the resonance
mass and $T^*$ (between 70 and 120 MeV) is a typical slope parameter for the
corresponding beam energy).  Many of the points are known and trivial, and can
trivially be implemented.  They are just of kinematical origin. However the
associated defects are by no means minor. Rather they ignore essential
features of the dynamics with consequences already on the {\em qualitative}
level and affect the spectra by far more than any of the currently
discussed im-medium effects, e.g. of the $\rho$-meson.  

\noindent
\begin{small}
\begin{tabular}{lll}
\multicolumn{2}{l}{\normalsize\bf List of defects in some of the transport
  codes} &{\normalsize\bf restoring factor}\\[2mm]
  [\makebox[2.5mm]{a}]&\parbox[t]{13.cm}{The differential mass information
  contained in the distribution functions $f(X,p)=f(X,m,{\vec p})$ of
  resonances is ignored and only the integrated total number is evaluated as a
  function of space-time (direct $\rho$ in
  refs. \cite{CassingpiA,Cassing}).\\[-2mm]}
  &\parbox[t]{3.4cm}{$\exp(-(m-m_R)/T^*)$\\(factor 10 or more at $m=500$ MeV
  for $\rho$)}\\[3mm] [\makebox[2.5mm]{b}]&\parbox[t]{13.cm}{Except for the
  $\pi^+\pi^-\rightarrow {\rm e}^+{\rm e}^-$ case most resonance production
  cross sections are parametrized such that they vanishes for $\sqrt{s}$
  values below the nominal threshold, e.g. below $m_N+m_{\rho}$ in the case
  $\pi N\rightarrow \rho N$. This violates detailed balance, since broad
  resonances can decay for $m<m_R$.\\[-2mm]} &\parbox[t]{3.2cm}{misses yield
  for\\ $m<m_R$}\\[3mm] [\makebox[2.5mm]{c}]&\parbox[t]{13.cm}{In {\em partial}
  cross sections leading to a resonance the randomly chosen mass is normally
  selected according to the spectral function. This is not correct since the
  corresponding {\em partial} width in the numerator of (\ref{A2}) has to be
  considered.\\[-2mm]} &\parbox[t]{3.2cm}{changes shape}\\[3mm]
  [\makebox[2.5mm]{d}]&\parbox[t]{13.cm}{Different partial cross sections are
  simply added without adjusting the total width in the resonance propagator
  accordingly. This violates unitarity.\\[-2mm]}
  &\parbox[t]{3.2cm}{$\left(\Gamma_{\rm free}/\Gamma_{rm tot}\right)^2$\\ at
  $m=m_R$}\\[3mm] [\makebox[2.5mm]{e}]&\parbox[t]{13.cm}{The Monte Carlo
  implementation of selecting the random mass $m$ of the \mbox{resonance}
  (item [c]) is sometimes falsely implemented, namely ignoring the kinetic
  phase-space of genuine multi-particle final state configurations, e.g. in
  $\pi N\rightarrow\rho N$.  Applies also to the $\Delta$-resonance , e.g. for
  $NN\rightarrow N\Delta$.\\[-2mm]} &\parbox[t]{3.2cm}{proportional to\\
  $\left({s}({\sqrt{s}-m-m_N})\right)^{1/2}$\\for two-body final state}\\[3mm]
  [\makebox[2.5mm]{f}]&\parbox[t]{13.cm}{For the electromagnetic decay of
  vector mesons some authors use a mass independent decay rate,
  e.g. $\Gamma_{\rho\rightarrow {\rm e}^+{\rm e}^-}/m={\rm const.}$, rather
  than that resulting from vector dominance and QED with
  $\Gamma_{\rho\rightarrow {\rm e}^+{\rm e}^-}\propto 1/m^2$.\\[-2mm]}
  &$(m_R/m)^3$
\end{tabular}
\end{small}

\section{$\Phi$-derivable approximations}

The preceding section has shown that one needs a transport scheme adapted for
broad resonances. Besides the conservation laws it should comply with
requirements of unitarity and detailed balance. A practical suggestion has
been given in ref. \cite{DB} in terms of cross section prescriptions. However
this picture is tied to the concept of asymptotic states and therefore not
well suited for the general case, in particular if more than one channel feeds
into a broad resonance. Therefore we suggest to revive the so-called
$\Phi$-derivable scheme, originally proposed by Baym \cite{Baym} on the basis
of a formulation of the generating functional or partition sum given by
Luttinger, Ward \cite{Luttinger}, and later reformulated in terms of
path-integrals \cite{Cornwall}. This functional can be generalized to
the real time case (for details see \cite{IKV1}) with the diagrammatic
representation\footnote{ $n_\Se$ counts the number of self-energy
  $\Se$-insertions in the ring diagrams, while for the closed diagram of
  $\Phi$ the value $n_\lambda$ counts the number of vertices building up the
  functional $\Phi$.}  \unitlength=.7cm
%
\begin{eqnarray}\label{keediag}
\hspace*{-0.8cm}
\ii\Gamma\left\{\Gr\right\} = \ii
\Gamma^0\left\{\Gr^0\right\} 
+
\underbrace{\sum_{n_\Se}\vhight{1.6}\frac{1}{n_\Se}\GlnG0Sa}
_{\displaystyle \pm \ln\left(1-\odot\Gr^{0}\odot\Se\right)}
\;\underbrace{-\vhight{1.6}\GGaSa}
_{\displaystyle \pm \odot\Gr\odot\Se\vphantom{\left(\Ga^{0}\right)}}
\;\;+\;\underbrace{\vhight{1.6}\sum_{n_\lambda}\frac{1}{n_\lambda}
\Dclosed{c2}{\thicklines}}
_{\displaystyle\vphantom{\left(\Ga^{0}\right)} 
+\ii\Phi\left\{\Gr\right\}}.
\end{eqnarray}
%
Thereby the key quantity is the auxiliary functional $\Phi$ given by
two-particle irreducible vacuum diagrams. It solely depends on fully
re-summed, i.e. self consistently generated propagators $\Gr(x,y)$ (thick
lines). The consistency is provided by the fact that $\Phi$ is the generating
functional for the re-summed self-energy $\Se(x,y)$ via functional variation
of $\Phi$ with respect to any propagator $\Gr(y,x)$, i.e.
\begin{eqnarray}\label{varphi}
-\ii \Se =\mp \delta \ii \Phi / \delta \ii \Gr. 
\end{eqnarray}
The Dyson equations of motion directly follow from
the stationarity condition of $\Gamma$ (\ref{keediag}) with respect to
variations of $\Gr$ on the contour\footnote{an extension to include classical
fields or condensates into the scheme is presented in ref. \cite{IKV1}}
%
\begin{eqnarray}
\label{varG/phi}
\delta \Gamma \left\{\Gr \right\}/ \delta \Gr = 0,
\quad&&\mbox{(Dyson eq.)}
\end{eqnarray}
%
In graphical terms, the variation (\ref{varphi}) with respect to $\Gr$ is
realized by opening a propagator line in all diagrams of $\Phi$.  The
resulting set of thus opened diagrams must then be that of proper skeleton
diagrams of $\Se$ in terms of {\em full propagators}, i.e.  void of any
self-energy insertions. As a consequence, the $\Phi$-diagrams have to be {\em
two-particle irreducible} (label $c2$), i.e. they cannot be decomposed into
two pieces by cutting two propagator lines.

The clue is that truncating the auxiliary functional $\Phi$ to a limited
subset of diagrams leads to a self consistent, i.e closed, approximation
scheme. Thereby the approximate forms of $\Phi^{\scr{(appr.)}}$ define {\em
  effective} theories, where $\Phi^{\scr{(appr.)}}$ serves as a generating
functional for the approximate self-energies $\Sa^{\scr{(appr.)}}(x,y)$
through relation (\ref{varphi}), which then enter as driving terms for the
Dyson equations of the different species in the system.  As Baym \cite{Baym}
has shown such a $\Phi$-derivable approximation is conserving for all
conservation laws related to the global symmetries of the original theory and
at the same time thermodynamically consistent. The latter automatically
implies correct detailed balance relations between the various transport
processes. For multicomponent systems it leads to a {\em actio} = {\em
  reactio} principle. This implies that the properties of one species are not
changed by the interaction with other species without affecting the properties
of the latter ones, too. The $\Phi$-derivable scheme offers a natural and
consistent way to account for this principle. Some thermodynamic examples have
been considered recently, e.g., for the interacting $\pi N \Delta$ system
\cite{Weinhold} and for a relativistic QED plasma \cite{Baym98}.

\section{Generalized Kinetic Equation}\label{sect-Kin-EqT}

In terms of the kinetic notation (\ref{F}) and in the first
gradient approximation the {\em generalized kinetic} equation for $F$
takes the form
%
\begin{equation}
\label{keqk}
\Do \F (X,p) =
B_{\rm in}(X,p) 
+ C (X,p) 
\end{equation}
%
with the drift term determined from
the ''mass'' function (c.f. (\ref{Aeq}))
%
\begin{eqnarray}\label{meqx}\label{M}
M(X,p)=M_0(p) -\Re\Se^R (X,p) 
\end{eqnarray}
through the Poisson bracket ${\Do F\equiv\Pbr{M,F}}$.
The explicit form of the differential drift operator reads
%
\begin{eqnarray}\label{Drift-O}
\Do = 
\left(
\vu_{\mu} - 
\frac{\partial \Re\Sa^R}{\partial p^{\mu}} 
\right) 
\partial^{\mu}_X + 
\frac{\partial \Re\Sa^R}{\partial X^{\mu}}  
\frac{\partial }{\partial p_{\mu}}
,  \quad\quad\mbox{with}\quad v^\mu=\frac{\partial M_0(p)}{\partial p_{\mu}}
=\left\{
\begin{array}{ll}
(1,{{\vec p}/m})\quad&\mbox{non-rel.}\\
2p^{\mu}&\mbox{rel. bosons.}
\end{array}\right.
\end{eqnarray}
The two other terms in (\ref{keqk}), $B_{\rm in}(X,p)$ and $C(X,p)$, are a
fluctuation term and the collision term, respectively
%
\begin{eqnarray}
\label{Coll(kin)}
B_{\rm in}=\Pbr{\Ldt , \Re\Gr^R},  \quad\quad
C (X,p) =
\Ldt (X,p) \Ft (X,p) 
- \Ld (X,p) \F (X,p),
.  
\end{eqnarray}
%
Here the reduced gain and loss rates and total width of the collision
integral are
%
\begin{eqnarray}
\label{gain}
\Ldt (X,p) &=&  \mp \ii \Se^{-+} (X,p),\quad\quad
\Ld (X,p)  =  \ii \Se^{+-} (X,p),\\ 
\label{G-def}
\Gamma (X,p)&\equiv& -2\Im \Se^R (X,p) = \Ld (X,p)\pm\Ldt (X,p). 
\end{eqnarray}
%
The combination opposite to (\ref{G-def}) determines the fluctuations
%
\begin{eqnarray}
\label{Fluc-def}
I (X,p) = \Ldt (X,p)\mp\Ld (X,p). 
\end{eqnarray}
%

We need still one more equation, which in fact can be provided by the retarded
Dyson equation. In first order gradient approximation the latter is completely
solved algebraically \cite{BM}
%
\begin{eqnarray}
\label{Asol}\label{Xsol}
&&\Gr^R=\frac{1}{M(X,p)+\ii\Gamma(X,p)/2}
\quad\Rightarrow\quad
A (X,p) =
\frac{\Gamma (X,p)}{M^2 (X,p) + \Gamma^2 (X,p) /4}
\end{eqnarray}
%
Canonical equal-time (anti) commutation relations
for (fermionic) bosonic field operators provide the standard sum--rule
for the spectral function.

We now provide a physical interpretation of the various terms in the
generalized kinetic equation (\ref{keqk}).  The drift term $\Do \Fd$ on the
l.h.s. of eq.  (\ref{keqk}) is the usual kinetic drift term including the
corrections from the self-consistent field $\Re\Se^R$ into the convective
transfer of real and also virtual particles.  For the collision-less case
$C=B=0$, i.e. \mbox{$\Do \Fd=0$} (Vlasov equation), the quasi-linear first
order differential operator $\Do$ defines characteristic curves. They are the
standard classical paths in the Vlasov case. Thereby the four-phase-space
probability $\Fd(X,p)$ is conserved along these paths. The formulation in
terms of a Poisson bracket in four dimensions implies a generalized Liouville
theorem. For the collisional case both, the collision term $C$ and the
fluctuation term $B$ change the phase-space probabilities of the
``generalized'' particles during their propagation along the ``generalized''
classical paths given by $\Do$. We use the term ``generalized'' in order to
emphasize that particles are no longer bound to their mass-shell, $M=0$,
during propagation due to the collision term, i.e.  due decay, creation or
scattering processes.

The r.h.s. of eq. (\ref{keqk}) specifies the collision term $C$ in terms of
gain and loss terms, which also can account for multi-particle processes.
Since $\Fd$ includes a factor $A$, the $C$ term further deviates from the
standard Boltzmann-type form in as much that it is multiplied by the spectral
function $A$, which accounts for the finite width of the particles.

The additional Poisson-bracket term
%
\begin{eqnarray}
\label{backflow}  
B_{\rm in}&=&\Pbr{\Ldt,\Re\Gr^R}=\frac{M^2-\Gamma^2/4}{(M^2+\Gamma^2/4)^2}\;
   \Do\;\Ldt
   +\frac{M\Gamma}{(M^2+\Gamma^2/4)^2}\Pbr{\Ldt,\Ld}
\end{eqnarray}
%
is special. It contains genuine contributions from the finite mass width of
the particles and describes the response of the surrounding matter due to
fluctuations. This can be seen from the conservation laws discussed below. In
particular the first term in (\ref{backflow}) gives rise to a back-flow
component of the surrounding matter. It restores the Noether currents to be
conserved rather than the intuitively expected sum of convective currents
arising from the convective $\Do\F$ terms in (\ref{keqk}).  The second term
of (\ref{backflow}) gives no contribution in the quasi-particle limit of small
damping width limit and represents a specific off mass-shell response due to
fluctuations, c.f.  \cite{LipS,IKV2}. In the low density and quasi-particle
limit the $B_{\rm in}$ term provides the virial corrections to the Boltzmann
collision term \cite{Mor98}.

\section{Conservations of the Current and Energy--Momentum}
\label{Conservation-L}

The global symmetries of $\Phi$ provide conservation laws such as the
conservation of charge and energy--momentum. The corresponding Noether
charge current and Noether energy--momentum tensor result to the following 
expressions, c.f. \cite{IKV1},
%
\begin{eqnarray}
\label{c-new-currentk}\nonumber 
j^{\mu} (X) 
&=&\frac{e}{2}\mbox{Tr} \int \dpi{p}
\vu^{\mu} 
\left(\Fd (X,p) \mp \Fdt (X,p) \right),\hspace*{-1cm} \\
\label{E-M-new-tensork}
\Theta^{\mu\nu}(X)
&=&\frac{1}{2}\mbox{Tr} \int \dpi{p} 
\vu^{\mu} p^{\nu} 
\left(\Fd (X,p) \mp \Fdt (X,p) \right)
+ g^{\mu\nu}\left(
{\cal E}^{\scr{int}}(X)-{\cal E}^{\scr{pot}}(X)
\right).  
\end{eqnarray}
%
Here 
%
\begin{eqnarray}
\label{eps-int} 
{\cal E}^{\scr{int}}(X)=\left<-\Lint(X)\right>
=\left.\frac{\delta\Phi}{\delta\lambda(x)}\right|_{\lambda=1},\quad
\label{eps-potk}
{\cal E}^{\scr{pot}}
= \frac{1}{2}\mbox{Tr}
\int\dpi{p} \left[
\Re\Sa^R \left(\Fd\mp\Fdt\right)
+ \Re\Ga^R\left(\Gb\mp\Gbt\right)\right]\nonumber
\end{eqnarray}
%
are the densities of the interaction energy and the potential energy,
respectively. The first term of ${\cal E}^{\scr{pot}}$ complies with
quasi-particle expectations, namely mean potential times density, the second
term displays the role of fluctuations $I=\Gb\mp\Gbt$ in the potential energy
density.  This fluctuation term precisely arises form the $B$-term in the
kinetic eq. (\ref{keqk}), discussed around eq. (\ref{backflow}). It restores
that the Noether expressions (\ref{E-M-new-tensork}) are indeed the exactly
conserved quantities. In this compensation we see the essential role of the
fluctuation term in the generalized kinetic equation. Dropping or
approximating this term would spoil the conservation laws. Indeed, both
expressions in (\ref{E-M-new-tensork}) comply exactly with the generalized
kinetic equation (\ref{keqk}), i.e. they are exact integrals of the
generalized kinetic equations of motion within the $\Phi$-derivable scheme.
Memory effects and the formulation of a kinetic entropy can likewise be
addressed \cite{IKV2}.\\ 

\noindent
{\bf Acknowledgement:} Much of the material presented is due a very
stimulating collaboration with Y. Ivanov and D. Voskresensky. The author
further acknowledges encouraging discussions with P. Danielewicz, B. Friman,
 H. van Hees, E. Kolomeitsev, M. Lutz, K. Redlich and W. Weinhold on various
aspects of broad resonances and J. Aichelin, S. Bass, E. Bratkowskaya,
W. Cassing, C.M. Ko and U. Mosel on some particular features of transport
codes.


\begin{thebibliography}{99}\itemsep-1.8mm
\bibitem{SKBK} J. Schwinger, {J. Math. Phys,} {\bf 2} (1961) 407;
L. P. Kadanoff and G. Baym, Quantum Statistical Mechanics (Benjamin,
1962); L. M. Keldysh, {ZhETF} {\bf 47} (1964) 1515; in
Engl. translation Sov. Phys. JETP{\bf 20 } (1965) 1018.
\bibitem{D}
P. Danielewicz, {Ann. Phys.} (N. Y.) {\bf 152} (1984) 239
\bibitem{Landsmann}N. P. Landsmann, Phys. Rev. Lett. {\bf 60} (1988)
1990; Ann. Phys. {\bf 186} (1988) 141.
\bibitem{DB}P. Danielewicz, G. Bertsch, Nucl. Phys. {\bf A533} (1991) 712.
\bibitem{BM}
W. Botermans and R. Malfliet, Phys. Rep. {\bf 198} (1990) 115.
\bibitem{HFN}M. Herrmann, B. L. Friman, W. N\"orenberg,
Nucl. Phys. {\bf A560} (1993) 411.
\bibitem{PH} P. A. Henning, Phys. Rep. {\bf C 253} (1995) 235;
Nucl. Phys. {\bf A 582} (1995) 633.
\bibitem{QH} E. Quack, P. A. Henning, GSI-95-29; Phys. Rev. Lett. in
print; GSI-95-42.
\bibitem{Weinhold} W. Weinhold, Diploma thesis, GSI 1995; W. Weinhold,
B. L. Friman, W. N\"orenberg, Acta Phys. Pol. {\bf 27}, Phys. Lett. {\bf B
433} (1998) 236.
\bibitem{KV} J. Knoll and D. N. Voskresensky, Ann. Phys {\bf 249} (1996)
532;\\ a condensed account of this work is published in
{Phys. Lett.}  {\bf B 351} (1995) 43.
\bibitem{Baym}
G. Baym, Phys. Rev. {\bf 127} (1962) 1391.
\bibitem{CGreiner}C. Greiner, K. Wagner, P.G. Reinhard,
  Phys.Rev. {\bf C49} (1994) 1693.
\bibitem{IKVTFT}J. Knoll et al., Proceedings of the Thermal Field Theory
  workshop, TFT 98, Regensburg 1998, hep-ph/9809419 
\bibitem{IKV2}Y. Ivanov, J. Knoll, D. Voskresenski, in preparation
\bibitem{LP} {E. M. Lifshitz and L. P. Pitaevskii}, ''Physical
Kinetics'' Nauka, 1979; Pergamon press, 1981.
\bibitem{IKV1}Yu. B. Ivanov, J. Knoll and D. N. Voskresenski,
  GSI-preprint-98-34, hep-ph/9807351.
\bibitem{BethU} E. Beth, G.E. Uhlenbeck, Physica 4 (1937) 915
\bibitem{Huang} K. Huang, "Statistical Mechanics", Wiley, New York (1963)

\bibitem{DMB} R. Dashen, S. Ma, H.J. Bernstein, Phys.Rev. 187 (1969) 345;
\bibitem{Mekjian} A.Z. Mekjian, Phys.Rev. C 17 (1978) 1051;
\bibitem{DP}P. Danielewicz and S. Pratt, Phys.Rev. {\bf C53} (1996) 249
\bibitem{Mor98}V. \v{S}pi\v{c}ka, P. Lipavsk\'{y}, K. Morawetz,
Phys. Lett. A{\bf 240} (1998) 160;
\bibitem{Weinh-PhD}W. Weinhold, Ph-D thesis, TU-Darmstadt, 1998
\bibitem{Mosel}S. Leupold, U. Mosel, Phys. Rev. {\bf C 58} (1998) 2939;
U. Mosel, these proceedings;
\bibitem{Rapp}R. Rapp, G. Chanfray, J. Wambach, Nucl. Phys. {\bf A 617}
(1997)472; 
\bibitem{Klingl}F. Klingl, N. Kaiser, W. Weise,  Nucl. Phys. {\bf A 624}
(1997) 527;
\bibitem{FrimanPirner}B.L. Friman, H.-J. Pirner, Nucl. Phys. {\bf A 617}
(1997) 496;
\bibitem{FLW} B. Friman, M. Lutz and G. Wolf, GSI-Preprint-98-63,
nucl-th/9811040; 
\bibitem{Ko}W.S. Chung, C.M. Ko, G.Q. Li, Nucl. Phys. {\bf A 641} (1998) 357;
\bibitem{CassingpiA}c.f.  Figs. 3 and 4 in W. Cassing et. al.,
Phys. Lett. {\bf B 396} (1997) 26 or Weidmann et al., nucl-th/9711004;
\bibitem{Cassing} c.f. Figs. 2 - 4 in W. Cassing et. al., Phys. Rev. {\bf C57}
(1997) 916, or
, and corresponding earlier references therein;
\bibitem{Luttinger} 
J. M. Luttinger and J. C. Ward, Phys. Rev. {\bf 118} (1960) 1417.
\bibitem{Cornwall} J.M. Cornwall, R. Jackiw and E. Tomboulis,
Phys. Rev.
{\bf D 10} (1974) 2428;
\bibitem{Baym98} B. Vanderheyden and G. Baym, hep-ph/9803300;
\bibitem{LipS}  V. Spicka and P. Lipavsky, Phys. Rev. Lett.
{\bf 73} (1994) 3439;  Phys. Rev. {\bf B52} (1995) 14615.
\end{thebibliography}
\end{document}